\documentstyle[12pt]{article}

\newcommand{\eq}{\begin{equation}}
\newcommand{\en}{\end{equation}}
\newcommand{\bea}{\begin{eqnarray}}
\newcommand{\eea}{\end{eqnarray}}
\newcommand{\spz}{\hspace{0.7cm}}

\newcommand{\virg}{\spz,\spz}

\newcommand{\resection}[1]{\setcounter{equation}{0}\section{#1}}

\newcommand{\G}{{\cal G}}

\newcommand{\NP}[1]{Nucl.\ Phys.\ {\bf #1}}
\newcommand{\PL}[1]{Phys.\ Lett.\ {\bf #1}}

\newcommand{\IJMP}[1]{Int.\ J.\ Mod.\ Phys.\ {\bf #1}}

\newcommand{\TMP}[1]{Teor.\ Math.\ Phys.\ {\bf #1}}

\begin{document}

\newcommand{\sps}{\hspace{3mm}}
\newcommand{\sts}{\footnotesize}
\newcommand{\stl}{\small}
\newcommand{\ba}{\begin{array}}
\newcommand{\ea}{\end{array}}
\setlength{\unitlength}{1mm}
\newsavebox{\pole}
\sbox{\pole}
{\begin{picture}(2,2)(0,0)
\put(0,0){\line(1,3){1.5}}
\put(0,0){\line(-1,3){1.5}}
\put(0,4.5){\oval(3,3)[t]}
\end{picture}}
\newsavebox{\An}
\sbox{\An}{\begin{picture}(52,5)(0,-3.5)
\put(0,0){$A_n$ :}
\multiput(10,0)(10,0){5}{\circle*{1}}
\multiput(10,0)(10,0){2}{\line(1,0){10}}
\multiput(31,0)(1,0){9}{\circle*{.2}}
\put(40,0){\line(1,0){10}}
\put(10,-1){\makebox(0,0)[t]{{\protect\sts 1}}}
\put(20,-1){\makebox(0,0)[t]{{\protect\sts 2}}}
\put(30,-1){\makebox(0,0)[t]{{\protect\sts 3}}}
\put(40,-1){\makebox(0,0)[t]{{\protect\sts {\em n}--1}}}
\put(50,-1){\makebox(0,0)[t]{{\protect\sts {\em n}}}}
\end{picture}}
\newsavebox{\Dn}
\sbox{\Dn}{\begin{picture}(55,15)(0,-7.5)
\put(0,0){$D_n$ :}
\multiput(10,0)(10,0){4}{\circle*{1}}
\multiput(10,0)(10,0){2}{\line(1,0){10}}
\multiput(31,0)(1,0){9}{\circle*{.2}}
\put(40,0){\line(1,1){7.07}}
\put(40,0){\line(1,-1){7.07}}
\put(10,-1){\makebox(0,0)[t]{{\sts 1}}}
\put(20,-1){\makebox(0,0)[t]{{\sts 2}}}
\put(30,-1){\makebox(0,0)[t]{{\sts 3}}}
\put(37,-1){\makebox(0,0)[t]{{\sts {\em n}--2}}}
\put(47.07,7.07){\circle*{1}}
\put(49.07,7.07){\makebox(0,0)[b]{{\sts {\em n}}}}
\put(50,-7.07){\makebox(0,0)[b]{{\sts {\em n}--1}}}
\put(47.07,-7.07){\circle*{1}}\end{picture}}
\newsavebox{\En}
\sbox{\En}{\begin{picture}(70,15)(0,-5)
\put(0,0){$E_n$ :}
\multiput(10,0)(10,0){5}{\circle*{1}}
\multiput(10,0)(10,0){2}{\line(1,0){10}}
\multiput(31,0)(1,0){9}{\circle*{.2}}
\put(40,0){\line(1,0){10}}
\put(30,0){\line(0,1){10}}
\put(30,10){\circle*{1}}
\put(10,-1){\makebox(0,0)[t]{{\sts 1}}}
\put(20,-1){\makebox(0,0)[t]{{\sts 2}}}
\put(30,-1){\makebox(0,0)[t]{{\sts 3}}}
\put(40,-1){\makebox(0,0)[t]{{\sts{\em n}--2}}}
\put(50,-1){\makebox(0,0)[t]{{\sts{\em n}--1}}}
\put(32,10){\makebox(0,0){{\sts{\em n}}}}
\end{picture}}
\newsavebox{\Tn}
\sbox{\Tn}{\begin{picture}(55,10)(0,-3.5)
\put(0,0){$T_n$ :}
\multiput(10,0)(10,0){5}{\circle*{1}}
\multiput(10,0)(10,0){2}{\line(1,0){10}}
\multiput(31,0)(1,0){9}{\circle*{.2}}
\put(40,0){\line(1,0){10}}
\put(50,0){\usebox{\pole}}
\put(10,-1){\makebox(0,0)[t]{{\protect\sts 1}}}
\put(20,-1){\makebox(0,0)[t]{{\protect\sts 2}}}
\put(30,-1){\makebox(0,0)[t]{{\protect\sts 3}}}
\put(40,-1){\makebox(0,0)[t]{{\protect\sts {\em n}--1}}}
\put(50,-1){\makebox(0,0)[t]{{\protect\sts {\em n}}}}
\end{picture}}
\newsavebox{\de}
\sbox{\de}{\begin{picture}(70,15)(0,-5)
\put(0,0){$E_6$ :}
\multiput(10,0)(10,0){5}{\circle*{1}}
\multiput(10,0)(10,0){4}{\line(1,0){10}}
\put(30,0){\line(0,1){10}}
\put(30,10){\circle*{1}}
\put(10,-1){\makebox(0,0)[t]{{\sts $\frac{8}{7}$}}}
\put(20,-1){\makebox(0,0)[t]{{\sts $\frac{25}{14}$}}}
\put(30,-1){\makebox(0,0)[t]{{\sts $\frac{157}{70}$}}}
\put(40,-1){\makebox(0,0)[t]{{\sts $\frac{25}{14}$}}}
\put(50,-1){\makebox(0,0)[t]{{\sts $\frac{8}{7}$}}}
\put(33,10){\makebox(0,0){{\sts $\frac{39}{28}$}}}
\end{picture}}
\newsavebox{\pde}
\sbox{\pde}{\begin{picture}(70,15)(0,-5)
\put(0,0){$E_7$ :}
\multiput(10,0)(10,0){6}{\circle*{1}}
\multiput(10,0)(10,0){5}{\line(1,0){10}}
\put(30,0){\line(0,1){10}}
\put(30,10){\circle*{1}}
\put(10,-1){\makebox(0,0)[t]{{\sts $\frac{13}{10}$}}}
\put(20,-1){\makebox(0,0)[t]{{\sts $\frac{41}{20}$}}}
\put(30,-1){\makebox(0,0)[t]{{\sts $\frac{13}{5}$}}}
\put(40,-1){\makebox(0,0)[t]{{\sts $\frac{157}{70}$}}}
\put(50,-1){\makebox(0,0)[t]{{\sts $\frac{9}{5}$}}}
\put(60,-1){\makebox(0,0)[t]{{\sts $\frac{81}{70}$}}}
\put(33,10){\makebox(0,0){{\sts $\frac{49}{230}$}}}
\end{picture}}
\newsavebox{\ude}
\sbox{\ude}{\begin{picture}(70,15)(0,-5)
\put(0,0){$E_8$ :}
\multiput(10,0)(10,0){7}{\circle*{1}}
\multiput(10,0)(10,0){6}{\line(1,0){10}}
\put(30,0){\line(0,1){10}}
\put(30,10){\circle*{1}}
\put(10,-1){\makebox(0,0)[t]{{\sts $\frac{3}{2}$}}}
\put(20,-1){\makebox(0,0)[t]{{\sts $\frac{7}{3}$}}}
\put(30,-1){\makebox(0,0)[t]{{\sts $\frac{103}{35}$}}}
\put(40,-1){\makebox(0,0)[t]{{\sts $\frac{37}{14}$}}}
\put(50,-1){\makebox(0,0)[t]{{\sts $\frac{23}{10}$}}}
\put(60,-1){\makebox(0,0)[t]{{\sts $\frac{13}{7}$}}}
\put(70,-1){\makebox(0,0)[t]{{\sts $\frac{6}{5}$}}}
\put(33,10){\makebox(0,0){{\sts $\frac{19}{10}$}}}
\end{picture}}

\renewcommand{\thefootnote}{\fnsymbol{footnote}}

\newpage
\setcounter{page}{0}
\begin{flushright}
Bologna preprint DFUB-92-11\\
Torino preprint DFTT-31/92\\
July 1992
\end{flushright}
\vskip .3cm

\vskip .7cm
\begin{center}
{\Large{\bf DYNKIN TBA'S}}\\
\vskip 1.3cm
{\large F.\ Ravanini$^1$, R.\ Tateo$^2$ and A.\ Valleriani$^1$}
     \footnote{E-mail: ravanini@bologna.infn.it, tateo@torino.infn.it,
     valleriani@bologna.infn.it}\\
\vskip .7cm
{\em $^1$ I.N.F.N. - Sez. di Bologna, and Dip. di Fisica,\\
     Universit\`a di Bologna, Via Irnerio 46, I-40126 Bologna, Italy\\
\vskip .4cm
     $^2$ Dip. di Fisica Teorica, Universit\`a di Torino\\
     Via P.Giuria 1, I-10125 Torino, Italy}\\
\end{center}
\vskip .7cm

\renewcommand{\thefootnote}{\arabic{footnote}}
\setcounter{footnote}{0}

\begin{abstract}
\noindent
We prove a useful identity valid for all $ADE$ minimal
S-matrices, that clarifies the transformation of
the relative thermodynamic Bethe Ansatz (TBA) from its standard
form into the universal one proposed by Al.B.Zamolodchikov.
By considering the graph encoding of the system of
functional equations for the exponentials of the pseudoenergies, we show that
any such system having the same form as those for the $ADE$ TBA's, can
be encoded on $A,D,E,A/Z_2$ only.
This includes, besides the known $ADE$ diagonal scattering,
the set of all $SU(2)$ related {\em magnonic} TBA's. We explore this
class sistematically
and find some interesting new massive and massless RG flows.
The generalization to
classes related to higher rank algebras is briefly presented and an intriguing
relation with level-rank duality is signalled.
\end{abstract}

\newpage

\resection{Introduction}

In the recent years the understanding of the topological properties of the
Renormalization Group (RG) space in two dimensions has undergone a
remarkable progress. First of all, the discovery of the dissipative nature of
the RG flows (the celebrated {\em c-theorem}~\cite{cth}) has
given a great insight into the problem.
Then, for a large class of RG flows that show the property of integrability,
the proposal by A.Zamolodchikov~\cite{Zam-ising} for the conjecture of a
factorizable S-matrix corresponding to
a certain perturbation of a Conformal Field
Theory (CFT) by one of its relevant operators,
allows a lot of non-perturbative information to be extracted. To give evidence
that the conjectured S-matrix really describes the considered theory, one must
extract from it information on the ultraviolet (UV) limit. This can be done
using the procedure recently introduced in this context by
Al.B.Zamolodchikov~\cite{Al-Potts}
that goes under the name of {\em Thermodynamic Bethe Ansatz} (TBA)
and whose original formulation traces back to Yang and Yang~\cite{YY}.

The TBA can be presented as a set of coupled non-linear integral equations
driving the evolution of the Casimir energy of the theory on a cylinder
along the RG flow {\em exactly}
and {\em non-perturbatively}. In spite of their apparent complexity, they are
often numerically integrable without using very heavy computer resources, for
each point on the RG flow, and show the peculiar behaviour to be {\em
analytically} solvable in the UV and infrared (IR)
limits, thanks to transformations
leading to sum rules of the Rogers dilogarithm function.

The deduction of TBA equations directly
from the S-matrix is easy only when the
latter is a purely elastic (diagonal) one. In the most general case of
non-diagonal S-matrix, one is lead
to consider Higher level Bethe Ansatz to get
a TBA system out of it.This is a formidable task in many cases, and
an alternative strategy would be welcome. In a beautiful and stimulating
piece of work Al.Zamolodchikov~\cite{Al-PL} noticed
that the TBA system for purely elastic
scattering matrices related to $A,D,E$ Lie algebras (i.e. those previously
treated in~\cite{KM}), can be suitably
transformed in a form where it appears a
clear encoding on the Dynkin diagram of the related $A,D,E$ algebra.

Reversing the strategy, one can draw a
diagram, set up a TBA on it, and compute
formally the central charge of the CFT (with action, say, $S_{CFT}$)
describing the UV limit, as well as the conformal dimension of the perturbing
operator $\Phi$. It is then reasonable to conjecture that the theory described
by the action
\eq
S=S_{CFT}+\lambda\int d^2x \Phi(x)
\en
put on a cylinder, has a Casimir energy driven along the RG flow by the
proposed
TBA. At this point other information can be extracted to test further this
conjecture and to explore, for example, the IR limit of such a
theory. Following this very productive attitude, in some recent
papers~\cite{Al-RSOS,Al-TIM,Al-coset,Martins,wtba} a lot of RG flows have been
studied, and even some new discovered~\cite{FatAl}.

While in the diagonal TBA, energy terms in the equations, one for each particle
type, were naturally attached to {\em all} the nodes of the corresponding
Dynkin
diagram, in the non-diagonal case the structure of the encoding is a bit
different, as there are a quantity of nodes with no energy term attached. These
nodes correspond to what in Bethe Ansatz literature are called {\em magnons},
i.e. fictitious particles with no mass and no energy, whose unique task is to
exchange internal degrees of freedom ({\em colors}) between the physical
particles of the theory. We shall often refer to such TBA systems as {\em
magnonic}.

In the present paper we attach the
program of systematically exploring this large class of TBA systems.
In~\cite{wtba} it was realized that for a large set of models, the encoding of
TBA is natural on a certain kind of ``product'' of two Dynkin diagrams, one for
the physical particle structure, the other for the magnonic structure. In
particular, when there is only one mass in the spectrum (so that the
``physical'' Dynkin diagram is $A_1$), the magnonic TBA are encoded on a single
Dynkin diagram. To be more precise, in~\cite{wtba} only the case of $A_n$
magnonic diagrams was explored. One can ask if the TBA encoded on, say, $D_n$
or
$E_n$ diagrams have any meaning, and even further, if there are some other
class of graphs, not Dynkin diagrams, that can encode the magnonic structure of
TBA. In the present paper we give an answer to these questions for the case
where the ``physical''
Dynkin diagram is $A_1$. It turns out that the magnonic TBA
has some basic properties (the so called {\em Y-system}, see below) in common
with the set of diagonal TBA's encoded on $ADE$ diagrams.

For the latter, this
Y-system is generated thanks to a useful identity on S-matrices that we prove
in
sect.2 (after review of the needed formalism of~\cite{Dorey}). Then
in sect.3 we
use this identity to give a proof of the transformations that lead Al.
Zamolodchikov to his universal form of TBA~\cite{Al-PL}. We also
get the Y-system
(i.e. a system of functional equations to be satisfied by the solutions of
TBA).
Sect.4 is devoted to the proof that a Y-system of the kind found in sect.3, can
be encoded {\em only} on $ADE$
Dynkin diagrams (plus the tadpole diagram $T_n$ corresponding
to a folding of $A_{2n}$). We show that a class of magnonic TBA's
generalizing those proposed in~\cite{Al-RSOS,Al-TIM,Al-coset,FatAl} has a
Y-system that simply maps into the previous one, thus allowing to extend the
$ADET$ classification to this case. Next, in sect.5, we explore systematically
the whole set of TBA's thus proposed, trying to identify their UV limit, the
perturbing operator, and, when applicable, the non-trivial IR limit. While
many of the
flows thus described were already known, some new appear, especially in the
study of the $E_{6,7,8}$ and $T_n$ cases. We end in sect.6 by commenting on the
generalizations when the ``physical'' diagram is not $A_1$, on the possibility
to envisage a general scheme for all TBA's and putting a remark on a still
mysterious relation with level-rank duality in CFT.

\resection{$ADE$ S-matrices: a useful identity}
We briefly summarize some basic facts about purely elastic scattering
theories which will be useful in the following. For our purposes it is
convenient to start from  Dorey's approach to the ADE S-matrices~\cite{Dorey}.
A $(1+1)$ dimensional purely elastic scattering theory has a
factorizable and diagonal S-matrix.
Factorizability means that the
scattering amplitudes of any number of particles
can be written as products of two-particle amplitudes.
Therefore,
the scattering of particles $a$ and $b$ is described by the two-particle
scattering amplitude $S_{ab}$, which is a function of the relative rapidity
$\theta_{ab}=\mid\theta_a-\theta_b\mid$. A simple pole of $S_{ab}$ at
$\theta_{ab}=i U^c_{ab}$ in the direct channel indicates that there exists
a bound state $c$ of $a$ and $b$ whose mass is:
\eq
m^2_c=m^2_a+m^2_b+2~m_a~m_b \cos(U^c_{ab})
\en
and the scattering amplitudes with any
particle $d$ must satisfy the bootstrap
equation:
\eq
S_{cd}(\theta)=S_{ad}(\theta+i \bar{U}^{\bar{b}}_{a \bar{c}})
S_{bd}(\theta-i \bar{U} ^{\bar{a}}_{b \bar{c}})
\label{boot}
\en
where $\bar{U}^{c}_{ab}=\pi-U^c_{ab}$. For the conserved charges, the
bootstrap equation leads to
\eq
 e^{-is \bar{U}^{\bar{b}}_{a \bar{c}} }q_s^a + e^{is
\bar{U}^{\bar{a}}_{b \bar{c}}}q_s^b=q_s^c
\en
In the so called $ADE$ scattering theories the fusing angles $U$ are all
integer multiples of $\pi \over h$, $h$ being the
Coxeter number of the $\G=ADE$ Lie algebra (of rank $r$)
associated to the theory.
It turns out that nontrivial solutions to the conserved charge
bootstrap only occur if the spin
$s$ modulo $h$, is equal to an exponent of $\G$.
Furthermore, each of the $r$ particles in the theory
may be assigned to a node on
the  Dynkin diagram of $\G$, in such a way that the set of conserved
charges of spin $s$, when
assembled into a vector $q_s=(q_s^{\alpha_1},
q_s^{\alpha_2},..q_s^{\alpha_r})$,
forms an eigenvector of the incidence matrix $G$
of the Dynkin diagram of $\G$
 with eigenvalue $2\cos({\pi s \over h})$. Thus,
\eq
G q_s=\lambda_s q_s \virg \lambda_s = 2\cos({\theta_s})
\virg \theta_s={\pi s \over h}
\label{g1}
\en
Notice that for $s=1$ eq.(\ref{g1}) gives the masses of the particles in
the theory, thus showing that they are organized in the so called
Perron-Frobenius eigenvector of $G$, namely in the eigenvector $\psi_G$
corresponding to its highest eigenvalue. For a matrix with
non-negative integer entries like $G$, $\psi_G$ turns out to be
always unique and has all non-negative components.

Let $\Phi$ be the root system of $\G$. Its Weyl group, i.e. the group of all
reflections
\eq
w_\alpha(x)=x-2 {(\alpha,x) \over (\alpha,\alpha)}\alpha \virg \alpha\in\Phi
\en
that map $\Phi$ into itself, is generated by the subset of the reflections
associated with the  set of simple roots $\Pi=\{\alpha_1,...\alpha_r\}$.
The Coxeter elements of the Weyl group  are elements of the form
$w_{\alpha_1} w_{\alpha_2}\cdots w_{\alpha_r}$.
Splitting $\Pi$ into two subsets of
orthogonal roots:
\eq
\Pi=({\alpha_1,\alpha_2 \cdots \alpha_k})\cup ({\beta_1,\beta_2 \cdots
\beta_{r-k}}) \virg
(\alpha_i,\alpha_j)=(\beta_i,\beta_j)=2 \delta_{i,j}
\en
one defines
\eq
w=w_{\alpha_1}w_{\alpha_2} \cdots w_{\alpha_k}
w_{\beta_1}w_{\beta_2} \cdots w_{\beta_{r-k}}
\en
$w$ is called a {\em Coxeter element} and has period $h$.
The group generated by
$w$, is therefore isomorphic to $Z_h$.
Let $\{\hat{\alpha},\hat{\beta}\}$ be the
dual basis to the simple root
$\{\alpha,\beta\}$,
$\lambda_s=2 \cos(\theta_s)$
an eigenvalue of the incidence matrix $G$ not
equal to 0, and $q_s$ the corresponding eigenvector.
Defining
\eq
a_s=\sum q_s^{\alpha_i} \hat{\alpha_i}~~~~~~~~~~~
b_s=\sum q_s^{\alpha_i} \hat{\beta_i}
\label{xx}
\en
following the arguments of~\cite{Dorey}, one can see that, for the
simple roots
\eq
(\alpha_i,a_s)=q_s^{\alpha_i}~~~~~~
(\beta_i,a_s)=0~~~~~~
(\alpha_i,b_s)=0~~~~~~
(\beta_i,b_s)=q_s^{\beta_i}
\en
and we can define a projector $P_s$ into the
two-dimensional subspace spanned by $a_s$ and $b_s$
\eq
P_s(\alpha_i)=q_s^{\alpha_i} \hat{a}_s~~~~~~
P_s(-\beta_j)=q_s^{\beta_j} \hat{b}_s
\en
where $\{ \hat{a}_s,-\hat{b}_s\} $ are
dual to $\{a_s,b_s \}$ in that subspace.
Since $a_s$ and $b_s$ have equal magnitude, this implies
that the projections of
the simple roots have lengths
proportional to the components $q_s^i$ of the
eigenvectors of the incidence  matrix.
The Coxeter element
$w$ acts in each subspace  as a rotation by $2\theta_s$. Hence we have
(introducing a complex notation in each invariant subspace)
\eq
P_s(w^p\alpha_i)=q_s^{\alpha_i} e^{i (2p+1)\theta_s}
\label{p1}
\en
\eq
P_s(w^p (-\beta_j))=-q_s^{\beta_j}  e^{i (2p)\theta_s}
\label{p2}
\en
In this formulation the general expression for the $S_{ab}-$matrix element
in the $ADE$ scattering theories are:

\noindent
(a) Particles $a$ and $b$ of type $\alpha$
\eq
 S_{ab}= \prod_{p=0}^{h-1} \{2p+1\}_+^{(\alpha_a,w^p \alpha_b)}
\label{ss1}
\en
(b) Particles $a$  of type $\alpha$ , $b$ of type $\beta$
\eq
 S_{ab}= \prod_{p=0}^{h-1} \{2p\}_+^{(\alpha_a,w^p \alpha_b)}
\label{ss2}
\en
(c) Particles $a$ and $b$ of type $\beta$
\eq
 S_{ab}= \prod_{p=0}^{h-1} \{2p-1\}_+^{(\alpha_a,w^p \alpha_b)}
\label{ss3}
\en
where
\eq
\{x\}_+=(x-1)^+(x+1)^+
\en
\eq
(x)^+=\sinh \left({\theta \over 2} + {i \pi x \over 2h}\right)
\en
We use this formalism to prove a useful identity:
\eq
S_{ab}\left(\theta+\frac{i\pi}{h}\right)
S_{ab}\left(\theta-\frac{i\pi}{h}\right) = \prod_c S_{ac}(\theta)^{G_{bc}}
\virg \theta \not= 0
\label{s1}
\en
The proof goes as follows. By using eq. (\ref{p1},\ref{p2}) it is
possible to rewrite equation  (\ref{g1})
\eq
\sum_b G_{ab}q_s^b = e^{\frac{i\pi s}{h}}q_s^a + e^{\frac{-i\pi s}{h}}q_s^a
\en
as
\eq
\alpha_a + \omega^{-1}\alpha_a = -\sum_b G_{ab}\beta_b \virg
-\omega\beta_b - \beta_b = \sum_a G_{ba}\alpha_a
\label{hhh}
\en
Then we have for case (a):
\eq
S_{ab}\left(\theta+\frac{i\pi}{h}\right)
S_{ab}\left(\theta-\frac{i\pi}{h}\right) =
 \prod_{p=0}^{h-1} \{2p+1\}_+^{(\alpha_a,w^p \alpha_b)}
 \prod_{k =0}^{h-1} \{2k-1\}_+^{(\alpha_a,w^k \alpha_b)}
\en
or, after a rescaling $k \to p+1$ (and using the property
$\{2h+x\}_+=\{x\}_+$)
\eq
S_{ab}\left(\theta+\frac{i\pi}{h}\right)
S_{ab}\left(\theta-\frac{i\pi}{h}\right) =
 \prod_{p=0}^{h-1} \{2p+1\}_+^{(\alpha_a,w^p(-\beta_b-w\beta_b))}
\en
and using the identity (\ref{hhh})
\eq
S_{ab}\left(\theta+\frac{i\pi}{h}\right)
S_{ab}\left(\theta-\frac{i\pi}{h}\right) =
 \prod_{p=0}^{h-1} \{2p+1\}_+^{\sum_k G_{bk}(\alpha_a,w^p\alpha_k)}=
\prod_{c}S_{ac}(\theta)^{G_{bc}}
\label{jjj}
\en
The other cases (a) and (b) (\ref{s1})  go through the same  way,
and so (\ref{ss1}-\ref{ss3}) does indeed provide a set of functions which
obey the S-matrix eq (\ref{s1}).

For $\theta = 0$ identity (\ref{s1}) must be carefully treated, as it becomes
apparent by taking its logarithmic form
\eq
\log S_{ab}\left(\theta+\frac{i\pi}{h}\right) + \log
S_{ab}\left(\theta-\frac{i\pi}{h}\right) = \sum_c G_{bc} \log S_{ac}(\theta)
- 2i\pi \Theta(\theta) G_{ab}
\label{A}
\en
The term proportional to the step function
\eq
\Theta(x)=\lim_{\epsilon\to 0}\left[\frac{1}{2}+\frac{1}{\pi}\arctan {x \over
\epsilon} \right] =
\left\{ \begin{array}{lll}
0 & {\rm if} & x<0 \\
\frac{1}{2} & {\rm if} & x=0 \\
1 & {\rm if} & x>0
\end{array} \right.
\en
has to be introduced to take into account the correct prescription for the
multivalued function $\log x$.
According to (\ref{A}), formula (\ref{s1}) must
be corrected as follows (if we want to include the point $\theta=0$)
\eq
S_{ab}\left(\theta+\frac{i\pi}{h}\right)
S_{ab}\left(\theta-\frac{i\pi}{h}\right) = \prod_c S_{ac}(\theta)^{G_{bc}}
e^{-2i\pi G_{ab}\Theta(\theta)}
\label{s2}
\en
The corrective exponential term is 1 for all values $\theta \not= 0$, while
at $\theta=0$ corrects the r.h.s. of the identity to be compatible with the
fact
that the l.h.s. becomes for $\theta=0$ the unitarity constraint on the matrix
$S$, while the values $S(0)$ must reproduce the correct statistics of the
system. We shall appreciate the deepness of this corrective term in next
section, where we relate it to the TBA equations.

We would like to emphasize
that (\ref{s2}) often gives relations equivalent to some
of the bootstrap equations (\ref{boot}). The deep interrelation between our
identity and the bootstrap certainly needs more investigation.

\resection{Universal form of TBA}

In~\cite{Al-Potts} it has been proposed, to recover the information on the
ultraviolet (UV) limit of the theory defined by the matrix $S_{ab}$, to use
the so called {\em Thermodynamic Bethe Ansatz} (TBA), which is a set of
non-linear coupled integral equations driving {\em exactly} the Casimir energy
of the system (on a cylinder of circumference $R$)
along its Renormalization Group (RG) flow, thus allowing the
determination of the effective central charge $\tilde{c}=c-24\Delta_0$
of the UV theory,
where the lowest conformal dimension $\Delta_0$ is 0 for unitary theories (for
which then $\tilde{c}=c$), and negative for non-unitary theories.
Putting $\nu_a=m_a R \cosh \theta$ (the so called {\em energy term}),
the TBA system is a
set of equations in the unknowns $\varepsilon_a$ (often called {\em
pseudoenergies}), having the general form
\eq
\nu_a(\theta) = \varepsilon_a(\theta) + \frac{1}{2\pi} \sum_b
(\phi_{ab}*\log (1+e^{-\varepsilon_b}))(\theta)
\label{tba}
\en
where $\phi_{ab}=-i{d \over d\theta} \log S_{ab}$ and
the $*$ stands for the rapidity convolution
\eq
(A*B)(\theta)=\int_{-\infty}^{+\infty} d\theta'A(\theta-\theta')B(\theta')
\en
{}From the solutions to this system, the evolution of the vacuum energy
$E(R)=-\frac{\pi \tilde{c}(R)}{6R}$ along
the RG flow can be followed by use of the equation
\eq
\tilde{c}(R) = \frac{3}{\pi^2}\sum_a \int_{-\infty}^{+\infty}\nu_a(\theta)
\log(1+e^{-\varepsilon_a(\theta)}) d\theta
\en
which, in the $R\to 0$ limit, turns out to be expressible, after
some manipulation involving the derivative of
eq.(\ref{tba}) (see for example ref.~\cite{KM}), in terms of Rogers
Dilogarithm\footnote{see for example~\cite{Lewin} for a definition and
for properties.} sum rules
\eq
\tilde{c}=\tilde{c}(0)=\frac{6}{\pi^2}\sum_a L\left(\frac{1}{1+y_a}\right)
\en
with $y_a$ given by the solutions to the algebraic trascendental equation
\eq
y_a = \prod_b (1+1/y_b)^{N_{ab}} \virg N_{ab}=-\int_{-\infty}^{+\infty}
\frac{d\theta}{2\pi} \phi_{ab}(\theta)
\label{y}
\en
that can be deduced from eq.(\ref{tba}) in the limit $R\to 0$.

In a nice recent piece of work~\cite{Al-PL} Al.B.Zamolodchikov proposed a
transformation of TBA equations for $ADE$ diagonal scattering
showing in a clear form their relation of the set of integral equations to the
$ADE$ Dynkin diagrams. This transformation,
leading to what is now known as {\em
universal form} of TBA, is based on a remarkable matrix
identity quoted in~\cite{Al-PL}
\eq
\left(\delta_{ab}-\frac{1}{2\pi}\tilde{\phi}_{ab}\right)^{-1} =
\delta_{ab} - \frac{1}{2\cosh(\pi k/h)} G_{ab}
\en
where $\tilde{\phi}_{ab}(k)$ stands for the Fourier transform of
$\phi_{ab}(\theta)$
\eq
\tilde{\phi}_{ab}(k) = \int_{-\infty}^{+\infty} d\theta
\phi_{ab}(\theta)e^{ik\theta}
\label{D}
\en
As in~\cite{Al-PL} there is no explicit proof of this identity, we give
here a proof
based on our identity (\ref{A}). We shall also explain in some more detail how
the universal form of TBA can be deduced out of it,
and also how one can obtain
a system of functional equations which is
also given in~\cite{Al-PL} and is very
useful in order to extract further information on the TBA system.

First of all, let us derive Zamolodchikov's identity from (\ref{A}).
Take the derivative of eq.(\ref{A})
and define as usual $\phi_{ab}(\theta) = -i
\frac{d}{d\theta} \log S_{ab}(\theta)$
\eq
\phi_{ab} \left(\theta+\frac{i\pi}{h}\right) + \phi_{ab}
\left(\theta-\frac{i\pi}{h}\right) = \sum_c G_{bc} \phi_{ac}(\theta) - 2\pi
\delta(\theta) G_{ab}
\en
and Fourier transform this equation ($k$ is the momentum corresponding to
$\theta$)
\eq
2\cos \left(\frac{k \pi}{h}\right)
\tilde{\phi}_{ab}(k) = \sum_c G_{bc} \tilde{\phi}_{ac}(k) - 2\pi G_{ab}
\label{B}
\en
or
\eq
\tilde{\phi}_{ab}(k) =  -2 \pi \left[ G \left({2\cos \left(\frac{
\pi k}{h}\right) - G}\right)^{-1}\right]_{ab}
\label{C}
\en
This equation is trivially equivalent to the matrix identity (\ref{D}),
but form (\ref{B}) is even more useful for our purposes. Notice that
(\ref{C}) computed in $k=0$ recovers a well known identity~\cite{KM,Dorey2}
\eq
N=A(2-A)^{-1}
\en
and this helps to transform eq.(\ref{y}) into the more appealing form
\eq
y_a^2 = \prod_b (1+y_b)^{G_{ab}}
\label{y2}
\en

Fourier transform eq.(\ref{tba}), then multiply both sides by
$\delta_{ab}-\tilde{R}(k)G_{ab}$, where $\tilde{R}(k)=
\frac{1}{2\cosh(\pi k/h)}$,
and finally use (\ref{B}) to recast the TBA system in the following {\em
universal} form
\eq
\nu_a(\theta) = \varepsilon_a(\theta) + \frac{1}{2\pi} \sum_b
G_{ab}[\varphi_h *(\nu_b - log (1+e^{\varepsilon_b}))](\theta)
\label{n}
\en
In this form, the TBA is explicitly fixed once the diagram whose
incidence matrix is $G$ is given.
The universal kernel $\varphi_h$, which is (up to $2\pi$) the Fourier
antitransform of $\tilde{R}(k)$, depends only on the Coxeter number $h$
of $G$
\eq
\varphi_h(\theta)= { h \over 2 \cos( { h \theta \over 2})}
\en
Notice that in the $R\to 0$ limit eq.(\ref{y2}) is directly obtained instead
of (\ref{y}).

Now let us consider eq.(\ref{tba}) for $\theta \to \theta-i\pi/h$ and for
$\theta \to \theta+i\pi/h$. Summing up and subtracting  eq.(\ref{tba})
calculated in $\theta$ and multiplied by $G$  and using (\ref{B}) we get
\eq
\nu_a(\theta+i\pi/h)+\nu_a(\theta-i\pi/h) - G_{ab}\nu_b(\theta) \\
= \varepsilon_a(\theta+i\pi/h) + \varepsilon_a(\theta-i\pi/h) -
G_{ab}\varepsilon_b(\theta) - G_{ab} L_b
\en
finally using the identity (\ref{g1}) for $s=1$
\eq
Y_a\left(\theta-\frac{i\pi}{h}\right) Y_a\left(\theta+\frac{i\pi}{h}\right) =
\prod_b (1+Y_b(\theta))^{G_{ab}}
\label{Y}
\en
Where $Y_a=e^{\varepsilon_a(\theta)}$.
This system (that we call {\em Y-system} in the following) is extremely
important, as commented by many authors, as it seems to encode even more
information on the system than the usual TBA.

First of all notice that the stationary solutions of this system (i.e.
those who do not depend on $\theta$) are exactly the $y_a$ appearing in eq.
(\ref{y2}), which are the basic tools to extract the UV central
charge.

Moreover, as stressed
in~\cite{Al-PL}, the Y-systems encoded on Dynkin diagrams show a remarkable
periodicity
\eq
Y_a\left(\theta+Pi\pi \right)=Y_{\bar{a}}(\theta) \virg P=\frac{h+2}{h}
\label{period}
\en
where $\bar{a}$ represents the antiparticle of $a$ (see fig.1).

\begin{figure}
\begin{center}
\begin{picture}(130,80)(0,-20)
\put(0,65){\usebox{\An}}
\put(63,68){{\stl $\bar{a}=n+1-a$,\sps$a=1,\ldots,n$}}
\put(0,45){\usebox{\Dn}}\put(60,52)
{\( \stl \left\{
  \ba{ll}
{\rm for}~n~{\rm even} & \bar{a}=a,~a=1,\ldots,n\\
{\rm for}~$n$~{\rm odd}  & \left\{ \ba{l}
                                 \bar{a}=a,~a=1,\ldots,n-2\\
                                 \bar{n}=n-1\ea\right.\ea\right. \)}
\put(0,25){\usebox{\En}}\put(60,30){\( \stl \left\{ \ba{l}
                                    {\rm for}~n=6:~\bar{1}=5,~\bar{2}=4,
                                    ~\bar{3}=3,~\bar{6}=6\\
                                    {\rm for}~n=7,8:~\bar{a}=a,
                                     ~a=1,\ldots,n\\
                                       n\leq 8 \ea\right. \)}
\put(0,5) {\usebox{\Tn}}\put(63,8){{\stl $(=A_{2n}/Z_2)$,\sps\sps$\bar{a}=a$,
\sps $a=1,\ldots,n$}}
\put(0,-10){\parbox{130mm}{\caption{\protect {\stl
$A,D,E,T$ diagrams: the numbers show the labelling of the
different nodes. On the right the particle--antiparticle relations
between nodes are shown.}}}}
\end{picture}
\end{center}
\label{fig1}
\end{figure}
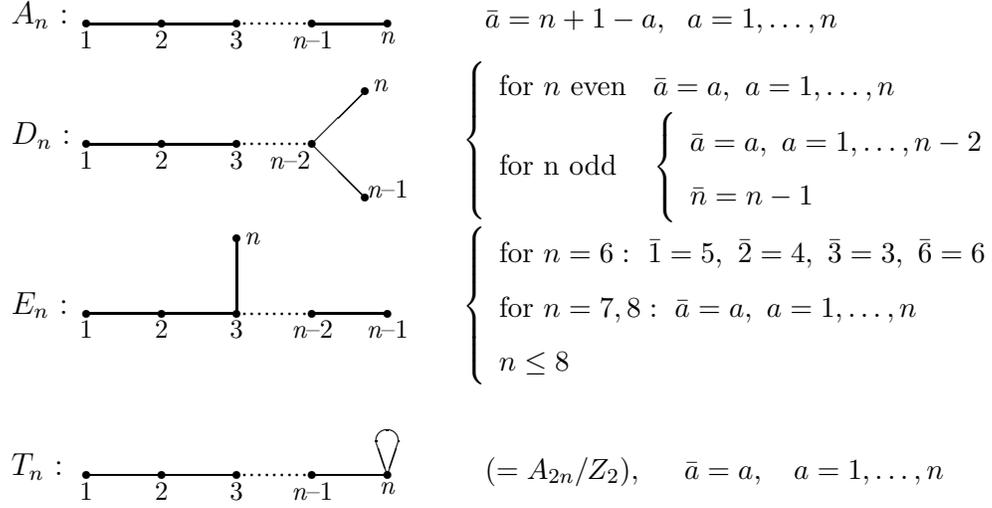

This can be shown (along the lines
of~\cite{Al-PL}) to be in relation with the
conformal dimension of the perturbing field, via the formula
\eq
\Delta=1-\frac{1}{P}~~{\rm for}~~A_n,D_n,E_n \virg
\Delta=1-\frac{2}{P}~~{\rm for}~~T_n
\label{delta}
\en
This allows to extract in a simple way
the parameter $\Delta$, characterizing,
together with the central charge $c$, the action of the theory. The explicit
proof of this periodicity relies on
successive substitutions inside the Y-system
of the functions $Y_a$ computed at different points. This
is in general a very
cumbersome task, even for the most simple cases.
We do not know of a general
proof of the periodicity, better we
used a simple computer program to test it up
to 16 digit precision for all Dynkin diagrams up to rank 50.

\resection{Y-systems on general graphs}

In the previous section we got the Y-system from $ADE$ massive scattering
theory. We can ask if such a system can be more generally defined on graphs
other than the Dynkin diagrams, thus allowing generalizations of the $ADE$
scattering theories. Here we prove that {\em any Y-system of the form
(\ref{Y}) where $G_{ab}$ is a matrix with non-negative integer entries, can
allow for stationary non-negative solutions ($y_a\geq 0$) only if $G_{ab}$
is the incidence matrix of an $A_n$, $D_n$, $E_{6,7,8}$ Dynkin diagram or the
``tadpole'' graph
$T_n=A_{2n}/Z_2$}. This set of graphs, shown in fig.1, defines the set of all
square
matrices with non-negative integer entries whose norm (the highest
eigenvalue, corresponding to the Perron-Frobenius eigenvector) is strictly
less than 2~\cite{GDJ}. The requirement of
existence of a stationary solution is a must to have a well defined central
charge for a system described by a TBA of form (\ref{n}).

As $\varepsilon_a$ are real functions, $e^{\varepsilon_a}=Y_a \geq 0$.
In particular, for solutions not depending on $\theta$, $y_a\geq 0$. More
precisely, it can never happen that $y_a=0$, otherwise in eq.(\ref{y2}),
at least one of the factors on the r.h.s. must be zero, implying one of the
$y_b$ to be $-1$, in contradiction with its positivity. Therefore we assume
$y_a > 0$ for all $a$.

Defining $x_a = \sqrt{y_a+1}$, eq.(\ref{y2}) becomes\footnote{The following
argument has been suggested to us by F.Gliozzi}
\eq
x_a^2 - 1 = \prod_b x_b^{G_{ab}}
\label{x}
\en
with $x_a>1$. Moreover, we can pose $z_a=\log x_a$ (and then $z_a>0$). This
allows to write the logarithm of the l.h.s. of (\ref{x}) as
\eq
\log (x_a^2-1) = 2z_a +\log\left(1-\frac{1}{x_a^2}\right) < 2z_a
\en
Therefore, from (\ref{x})
\eq
\sum_b G_{ab} z_b < 2z_a
\label{diseg}
\en
$z_a$ can be decomposed in the base of eigenvectors of $G$ and, having all
positive components, it has a positive and
non-zero projection on the Perron-Frobenius
eigenvector. A projection of formula (\ref{diseg}) on the Perron-Frobenius
direction simply shows that $\lambda_P<2$, where $\lambda_P$ is the
Perron-Frobenius eigenvalue. This bound is known to select the incidence
matrices of the graphs $A,D,E,T$ drawn in fig.1.

Corresponding to the Y-systems of the form (\ref{Y}) we have seen that
there are TBA systems of the (universal) form (\ref{n}). These are nothing
but the whole set of TBA's studied in the paper of Klassen and
Melzer~\cite{KM} (including the case they call $A_{2n}^{(2)}$ that
corresponds, in our notation, to the $T_n$ diagrams). We refer
to~\cite{KM} for a complete description of the identification of the
models at UV, and of their perturbing operators.

We notice, however, that the proof of classification of Y-systems
we have given is
absolutely independent of $h$. Other choices of the parameter $h$, where it no
more plays the role of Coxeter number, can, in principle,
lead to sensible TBA systems. One such choice, on which we shall concentrate
in the following, is the {\em magnonic} TBA proposed by Al.B.
Zamolodchikov~\cite{Al-RSOS,Al-TIM,Al-coset} to describe RG flows of
minimal models perturbed by their least relevant operator $\phi_{13}$. This
TBA has the general diagrammatic form
\eq
\nu_a(\theta) = \varepsilon_a(\theta) + \frac{1}{2\pi} \sum_b G_{ab}
(\phi * L_b) (\theta) \virg \phi(\theta)=\frac{1}{\cosh \theta}
\label{magn}
\en
The rationale under this form of TBA will appear later~\cite{wtba}. Here the
terms $\nu_a$ are zero on all nodes but one or two (labelled $k,l$ in the
following formula)
\eq
\nu_a(\theta)=\left\{ \begin{array}{ll}
\delta_a^k mR \cosh \theta & {\rm for~massive~case,}~\lambda<0 \\
\nu_a = \frac{mR}{2} (\delta_a^k e^{\theta} + \delta_a^l
e^{-\theta}) & {\rm for~massless~case,}~\lambda>0
\end{array} \right.
\label{nu-magn}
\en
Taking this equation for $\theta\to \theta-\frac{i\pi}{2}$, summing it
to the same equation for $\theta\to \theta+\frac{i\pi}{2}$ and taking into
account the pole structure of the kernel $\phi=1/\cosh\theta$ and the fact
that if $\nu_a$ are given by eq.(\ref{nu-magn}) then $\nu_a(\theta+i\pi/2)+
\nu_a(\theta-i\pi/2) = 0$, one arrives
at an equation having the same form of (\ref{Y}), where now $Y_a(\theta) =
e^{-\varepsilon_a(\theta)}$ (notice the different sign in the exponential).
The stationary solutions to this system again are the basic tool to compute the
UV central charge.

The surprising fact is that in this case the lines of reasoning that
lead to the $A,D,E,T$ classification of Y-system (\ref{Y}) apply as well,
therefore such a classification holds for this magnonic structures too.
Next section is devoted to the exploration of all possible such
magnonic systems on $A,D,E,T$.

\resection{$ADET$ magnonic TBA}

To begin this section, we would like to emphasize some general
rules on the diagrammatic approach to the magnonic TBA (\ref{magn}),
that we use in the
following. Then we briefly comment on non-perturbative terms, and finally
present our results on the systematic exploration of all $ADET$ cases.

\subsection{$c_{UV}$ in massive models and dilogarithm sum rules}

For massive models, imagine that the single massive term $\nu_a$ different
from zero is put on node $k$. Then to compute the UV central charge
consider the full diagram $G$ and the diagram $G'=G-\{ k\}$ where the node $k$
and the links emanating from it have been deleted. Then, following the
arguments in~\cite{Al-Potts,KM,Al-RSOS,Al-TIM,Al-coset}, we have the
following general rule
\eq
c_{UV} = \frac{6}{\pi^2} \left( \sum_{a\in G} - \sum_{a\in G'} \right)
L\left(\frac{y_a}{1+y_a}\right)
\label{cuv}
\en
$c_{IR}$ instead is zero, as the model is massive and has a trivial IR
point.

Formula
(\ref{cuv}) is easily computed by resorting to the dilogarithm sum rules
quoted in~\cite{KM} and remembering the following identity on Rogers
dilogarithm
\eq
L\left(\frac{x}{1+x}\right) = \frac{\pi^2}{6} - L\left(\frac{1}{1+x}\right)
\en

\subsection{$c_{UV}$, $c_{IR}$ and parity issues for massless models}

For massless models, call $R$ the node on which the left mover
$\nu=\frac{mR}{2} e^{\theta}$ is put, and $L$ the node on which the right
mover $\nu=\frac{mR}{2} e^{-\theta}$ lies. Then consider the diagrams
$G'=G-\{ L\}$ and $G''=G-\{ L,R\}$. The UV central charge is given by
\eq
c_{UV} = \frac{6}{\pi^2} \left( \sum_{a\in G} - \sum_{a\in G'} \right)
L\left(\frac{y_a}{1+y_a}\right)
\en
and therefore coincides with the calculation for the corresponding massive
case, where the mass is put on the same node as the left mover. The IR
central charge instead is given by
\eq
c_{IR} = \frac{6}{\pi^2} \left( \sum_{a\in G'} - \sum_{a\in G''} \right)
L\left(\frac{y_a}{1+y_a}\right)
\en

Notice that the position of the right and left movers, i.e. the
choice of nodes $L$ and $R$ is not arbitrary. Parity invariance of the
vacuum requires that the TBA system must be invariant under exchange
$\theta\to -\theta$ and therefore $R$ and $L$ must be interchangeable on
the diagram without changing the TBA structure, i.e. they must lie on nodes
symmetric with respect to some $Z_2$ symmetry of the diagram. This is a
strong constraint on the possible massless flows. For example $E_{7,8},T_n$
do not possess any $Z_2$ symmetry and it is not possible to write a
sensible TBA describing a massless flow on them. This is in connection with
the parity of the perturbing operator. If the operator is even, two
different behaviours are in general expected for different signs of the
perturbing parameter, one being massive and the other a massless crossover
to a non-trivial IR theory. Conversely, for parity odd operators the sign
of the perturbing parameter can always be readsorbed in the operator and
does not affect the (massive) behaviour of the perturbed theory.

\subsection{Periodicity of Y-system and conformal dimension of the perturbing
operator}

The conformal dimension $\Delta$
of the perturbing operator can be deduced from the periodicity of the
Y-system and is {\em independent} on the choice of the particular nodes
where masses or left-right movers are put. Of course, as the role of $h$ is
changed, the periodicity also gets some modification. Eq.(\ref{period}) still
holds, but now
\eq
P=\frac{h+2}{2}
\en
Moreover, the symmetry on the diagrams for $a\leftrightarrow \bar{a}$ is now
destroyed by the asymmetric choice for $\nu_a$. When $a \not= \bar{a}$ in
fig.1,
the real periodicity is doubled. One should be careful, however,
that this prediction can be affected by some selection rule on correlation
functions of the perturbing operator at criticality coming from
symmetries of the conformal fusion rules governing the UV CFT. A careful
analysis of all the cases leads anyway to a formula like (\ref{delta}). For
reader's convenience, we list here the results for all diagrams
\eq
\begin{array}{lllll}
A_n: & \Delta=1-\frac{2}{n+2}     & \spz & E_6: & \Delta=\frac{6}{7} \\
D_n: & \Delta=1-\frac{1}{n}       & \spz & E_7: & \Delta=\frac{9}{10} \\
T_n: & \Delta=1-\frac{2n-1}{2n+3} & \spz & E_8: & \Delta=\frac{15}{16}
\end{array}
\label{delta-adet}
\en

\subsection{Non-perturbative terms in the Casimir energy}

Once the UV and IR behaviours are identified, the expansion of the Casimir
energy in terms of $R$ is an issue. This contains both perturbative (in
$g=R^{2-2\Delta}$) and non perturbative terms.
The non-perturbative contributions
can be computed along the lines of~\cite{Al-Potts,KM,Al-RSOS,Al-TIM,Al-coset},
and amount or to a bulk term proportional to $R^2$ or to a logarithmic term
proportional to $R^2 \log R$. This latter appears when the incidence matrix $G$
is not invertible. In other words, the scale function $F(R)
=\frac{RE(R)}{2\pi}$, for $E_6,E_8,A_{2n}$ has the general form
\eq
F(R)= - {c \over 12} + {\epsilon_0 R^2 \over 2\pi} +
\sum_{n=1}^{\infty} f_n g^n
\en
while in the other cases $D_n,A_{2n+1},E_6$ the logarithmic bulk term appears
\eq
F(R)= - {c \over 12} +  L_k \log(mR) \left(\frac{mR}{2\pi}\right)^2
+ \sum_{n=1}^{\infty} f_n g^n
\en
where the coefficient $L_k$ pertains only the node $k$ where the mass (or the
left mover) is put, and can be elegantly expressed by considering the
eigenvector $q$ corresponding to the null eigenvalue, i.e.
\eq
\begin{array}{lllll}
A_{n~odd}: & q=(1,0,-1,0,1,\cdots) & \spz & E_7: & q=(1,0,0,0,0,-1,-1) \\
D_{n~odd}: & q=(0,0,\cdots,0,-1,1) & \spz & D_{n~even}: &
                                            q^{(1)}=(0,0,\cdots,-1,1) \\
           &                       & \spz &            &
                                            q^{(2)}=(\cdots,0,2,0,-2,0,1,1)
\end{array}
\en
and correspondingly
\eq
\begin{array}{lllll}
A_{n~odd}:    & L_k = -\frac{2}{n+3} (q_k)^2 & \spz &
E_7:          & L_k = -\frac{10}{3} (q_k)^2 \\
D_{n~odd}:    & L_k = -\frac{n-1}{2n}(q_k)^2 & \spz &
D_{n~even}:   & L_k = -\frac{n-1}{2n} (q_k^{(1)})^2 -\frac{1}{2n}(q_k^{(2)})^2
\end{array}
\en
\subsection{Model identifications}
In the following we summarize and comment the
results concerning our exploration of all
possible $ADET$ magnonic TBA structures. We divide the list according to the
Dynkin diagram, and for each Dynkin diagram we first put a massive energy
term $\nu_k=
mr\cosh\theta$, and let $k$ vary along the diagram up to exhaustion. Then,
if the diagram presents some $Z_2$ symmetry $k \leftrightarrow \bar{k}$,
we put $\nu_k = mRe^{\theta}/2$ and $\nu_{\bar{k}} =
mRe^{-\theta}/2$ and let $k$
vary on all nodes of the diagram with non trivial image under this $Z_2$.

\subsubsection{$A_n$ case}

The work concerning $A_n$ Dynkin diagrams has already
been done completely by Al.B.Zamolodchikov in the series of
works~\cite{Al-RSOS,Al-TIM,Al-coset}. For reader's convenience, we summarize
here his results. To take advantage of the $Z_2$ symmetry of the $A_n$ diagram,
it is convenient to put $n=k+l-1$ and consider $A_{k+l-1}$. The $\nu_a$'s are
chosen as in (\ref{nu-magn}).
The resulting flows start at the ($k,l$)-th $SU(2)$ coset model,
perturbed by its operator of dimension $\Delta=1-\frac{2}{k+l+1}$ and when
$\lambda<0$ evolve into the massive theory described by the non-diagonal
S-matrices of Ahn, Bernard, LeClair~\cite{ABL},
while for $\lambda>0$ the flow is
massless and at the IR limit reaches the ($k-l,l$)-th $SU(2)$-coset.

\subsubsection{$D_n$ case, mass on the tail}

Putting a mass term $\nu_k=mR\cosh\theta$ on the $k$-th node of the tail
($k=1,...,n-2$) of a $D_n$ diagram, one can describe a massive RG flow whose UV
limit has central charge
\eq
c=\frac{3k}{k+2}
\en
and therefore lies on the $k$-th critical line of those described
in~\cite{DSZ}\footnote{By considering the ground state TBA we can not
distinguish a model from its orbifold (this could be done by considering
excited
states TBA, instead), therefore here and in the following we ``identify''
models
differing one from the other by some orbifolding procedure. For example, we
speak of a single critical line at $c=1$, while it is known that a more subtle
analysis shows~\cite{curiosities} two lines which are one the orbifold of the
other.}.
This critical line, in turn, can be seen as the UV limit of a fractional
super-Sine-Gordon theory~\cite{LeClair},
and the perturbing operator can therefore be
identified with $\Phi\bar{\Phi}$ where
\eq
\Phi = \psi_1 :e^{\frac{i \beta}{\sqrt{4\pi}} \phi}:
\label{Phi}
\en
$\psi_1$ being the $Z_k$ generating parafermion of dimension $1-1/k$ and $\phi$
a free massless bosonic field, so that the vertex operator in (\ref{Phi}) has
dimension $\beta^2/8\pi$. The dimension of $\Phi$
must fit the value predicted by
the periodicity of the Y-system (\ref{delta-adet}).
Notice that the value of $c$ depends only on $k$ (the node where we put the
mass) and {\em not} on $n$, the rank of the diagram. Hence, for each $k$ there
are a sequence of points on the critical line labelled by $n$.
The identification is done by comparing the
dimension of the perturbing operator as predicted by the periodicity of the
Y-system, namely $1-1/n$ and the dimension of $\Phi$ as described above. This
yields
\eq
\frac{\beta^2}{8\pi} = \frac{1}{k} - \frac{1}{n}
\en
Notice that this result for $k=1$ was known to Al.Zamolodchikov, as quoted
in~\cite{FatAl}. This allows to identify the S-matrix of the perturbed massive
theory as
\eq
S=S_k \otimes S_{SG}\left(\frac{1}{k}-\frac{1}{n}\right)
\en
where $S_k$ stands for the Bernard LeClair~\cite{BL} $k$-th
minimal model + $\phi_{13}$ S-matrix and $S_{SG}(\beta^2/8\pi)$ means the
Sine-Gordon S-matrix~\cite{ZamZam-annals} at coupling $\beta$.

Notice that our TBA for such S-matrix is in agreement with the recent
observation in~\cite{FI} about the gluing (at the ``massive'' node)
of diagrams pertaining each factor in
a TBA corresponding to a tensor product S matrix.

Finally we would like to signal that when $k=n-2$, i.e. when the mass goes on
the bifurcation point of the diagram, we have the $N=2$ supersymmetric point of
the corresponding critical line, as stressed recently by Fendley and
Intriligator~\cite{FI}.

\subsubsection{$D_n$ case, mass on the fork}

The other possibility is to put the mass on the node $n$ (or equivalently
$n-1$). This case has been analyzed by Fateev and
Al.Zamolodchikov~\cite{FatAl}. We report it here for completeness.
The UV central charge calculation gives
\eq
c_{UV}=\frac{2(n-1)}{n+2}
\en
thus showing a dependence on $n$ in this case. This turns out to be the central
charge of the celebrated $Z_n$ parafermionic models, $SU(2)_n/U(1)$ as coset
models. The perturbation, as usual, is identified by the periodicity of the
Y-system to have $\Delta=1-\frac{1}{n}$, and it is therefore identified with
the
operator $\psi_1(z)\bar{\psi}_1(\bar{z})+\psi_1^{\dagger}(z)
\bar{\psi}_1^{\dagger}(\bar{z})$, where $\psi_1$ is the generating parafermion.
This perturbation is parity even, hence we expect it to be sensitive to the
sign
of the coupling $\lambda$.

Indeed the $D_n$ Dynkin diagram has an evident $Z_2$ symmetry exchanging $n$
with $n-1$ and allowing the definition of a massless TBA flowing in the
direction opposite to the previous one. In this case put a left mover on $n$
and
a right mover on $n-1$. The UV central charge is the same as before, but
the IR one is now given by
\eq
c_{IR}=1-\frac{6}{(n+1)(n+2)}
\en
thus giving evidence of a deeply non-perturbative flow between the
$Z_n$-parafermion model and the $n+1$-th minimal model. This result shows how
the approach of conjecturing a TBA and then trying to identify the UV and IR
limits by use of the Y-system is a very effective one: it can give evidence of
highly non-trivial and even unexpected results on the structure of the RG space
in two dimensions.

\subsubsection{$E_n$ case: mass on node 1}

This is perhaps the most intriguing case of our analysis. The three cases yield
the following values of the UV central charge and of the perturbing operator
conformal dimension
\eq
\begin{array}{lll}
E_6 & c=\frac{8}{7} & \Delta=\frac{6}{7} \\
E_7 & c=\frac{13}{10} & \Delta=\frac{9}{10} \\
E_8 & c=\frac{3}{2} & \Delta=\frac{15}{16}
\end{array}
\en
To identify the sequence of models giving the UV limit, it
is interesting to complete this table by extending the $E_n$ diagram to $n<6$
by
taking $E_5=D_5$, $E_4=A_4$, $E_3=A_2\oplus A_1$. The
second $A_1$ factor in the last case is a pure magnon decoupled from the
theory and it drops. Only the first factor is relevant. This allows to extend
the previous table with the additional cases
\eq
\begin{array}{lll}
E_3 & c=\frac{7}{10} & \Delta=\frac{3}{5} \\
E_4 & c=\frac{6}{7} & \Delta=\frac{5}{7} \\
E_5 & c=1 & \Delta=\frac{4}{5}
\end{array}
\en
What is peculiar with all these UV models is that they share the property to be
invariant under a generalized parafermionic algebra with $Z_K$ grading (i.e.
fusion rules $\psi_i \times \psi_j = \psi_{i+j~mod~K}$) and generating
parafermion $\psi_1$ of dimension $1+1/K$. These algebras have been called
$SZ_K$ in~\cite{exotic}.
Here $K$ is related to $n$ of $E_n$ by
$K=n-1$. It is expected that such theories are the UV limit of the S-matrices
proposed in~\cite{BP} having $Z_K$-exotic supersymmetry. The surprising fact
is that, as shown in~\cite{exotic} (see also~\cite{pfurlan}),
with the hypothesis that no operator of
dimension 1 appears as secondary of the identity,
this series of algebras truncates at $K\leq
6$. To be more precise, the $SZ_2$ algebra is the $N=1$ superconformal algebra,
generated by a field of spin $3/2$, then $SZ_3$ is the spin 4/3 algebra of
Fateev and Zamolodchikov~\cite{FatZam-s3},
the $SZ_4$ algebra describes a model on the
$c=1$ critical line invariant under a symmetry generated by a spin 5/4 field,
the two remaining cases are degenerate: $SZ_5 \equiv Z_5$ and $SZ_6 \equiv Z_2
\otimes Z_3$, where $Z_N$ are the usual parafermions. Notice that the central
charges of the models described by these algebras (for $K=2,3$ the bottom
models
of the relative series of minimal models) are exactly those arising in our TBA
computation. It is then tempting to identify the off-critical versions of these
models (perturbed by the operators indicated above) with the scattering
theories
having the Bernard Pasquier S-matrices. However a first question immediately
arises: what corresponds as UV to S-matrices with $K>6$? This is up to now an
open question. Secondly, we could for a while rejoice seeing that the truncated
series of $SZ_K$ corresponds to an $E_n$ series, which is truncated too.
However
the two truncation are not at the same $K$. The case of $E_8$ seems not to
enter
this framework. The value of 3/2 for the central charge for $E_8$ suggests that
we are dealing with a model at a specific point of the super-Sine-Gordon line.
If we try to identify this model by use of the perturbing operator dimension
(as before for the $D_n$'s) we get a theory with $\beta^2/8\pi=7/16$ that can
be
shown
to possess a parafermionic symmetry generated by a $\psi_1$ of dimension 8/7!
(now there is an operator of dimension 1 as a secondary of the identity, hence
the hypothesis of~\cite{exotic,pfurlan}
do not apply here). What are the generalizations
of this series for, say, parafermions of dimension 9/8, 10/9, etc...? This is
an
intriguing problem, and although its solution lies out of the scope of the
present paper, we intend to return on this point in future.

To conclude, we notice that the $E_{7,8}$ diagrams do not
possess any $Z_2$ symmetry and it is not possible to define massless flows on
them. However, for the $E_6$ diagram, we can transform the mass on the first
node to a left mover and put a right mover on the symmetric node $\bar{1}=5$.
This shows that the perturbation of the $Z_5$ model by its second energy
operator $\varepsilon_2$ of dimension 6/7 is even. For negative values of
the perturbing parameter it flows to a
massive scattering theory described by the
aforementioned Bernard-Pasquier S-matrix, while for $\lambda>0$ this defines a
massless theory flowing to an IR limit that can be easily be computed to have
$c=1$.

\subsubsection{$E_n$ case: mass on other nodes}

For $E_n$ and mass terms on nodes other than 1, the results of the
calculations of $c_{UV}$ have been encoded, for reader's
convenience, in fig.2.
We do not enter in much detail on the identification of UV models for these
cases. Most of them can be identified with tensor products of mutually
non-interacting minimal models.

\begin{figure}
\begin{center}
\begin{picture}(90,80)(-10,0) 
\put(0,75){\usebox{\de}}
\put(0,55){\usebox{\pde}}
\put(0,35){\usebox{\ude}}
\put(-10,16){\parbox{90mm}{\protect{
\caption{ \protect{\small
$E_{6,7,8}$ Dynkin diagrams: the numbers are the values of
the UV central charge
when the massive energy term is put on the corresponding node.}}}}}
\end{picture}
\end{center}
\label{fig2}
\end{figure}
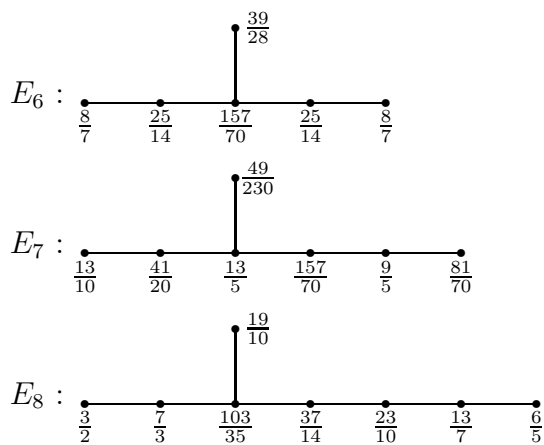

We discuss a single case which can have some interest by itself.
Putting a mass term on node 2 of the $E_6$ diagram we get
$c_{UV}=\frac{25}{14}$, which corresponds to the tensor product of two copies
of
the $m=7$ minimal model\footnote{The other possible identification with the
$Z_{26}$ parafermion is ruled out by the non-existence of an operator of
dimension 6/7 in its Kac-table}
The $\Delta=\frac{6}{7}$ perturbing field is realized as the tensor product
of operators of dimensions 3/28 and 3/4 respectively.
The perturbation happens to be parity even as one
can figure out from the known parity of operators in $m=7$ minimal model. This
is in agreement with the $Z_2$ symmetry of the $E_6$ diagram. Therefore,
by replacing
the mass on node 2 by a left mover and putting a right mover on node 4 we
recover a massless flow between the aforementioned model and an IR limit with
$c_{IR}=81/70$ that we can identify with the $m=5$ model of
the $N=1$ superconformal series.

\subsubsection{$T_n$ case: a new series of non-unitary massive flows}

Another quite unexpected result concerns the tadpole diagrams $T_n$. Here it is
convenient to introduce the parameter $p=2n+1=3,5,7,...$.
Notice first of all that the $T_n$
diagram has no $Z_2$ symmetry, so we can expect pure massive flows only, and
the
perturbing operator will be odd.
Put a mass term $\nu_l=mR\cosh\theta$ on node $l$ of $T_n$. The central charge
computation gives
\eq
\tilde{c} = \frac{3l}{l+2}\left(1-\frac{2(l+2)}{p(p+2l)}\right)
\en
The central charge computed here is, to be
precise, an effective one $\tilde{c}=c-24\Delta_0$, where
$\Delta_0$ is the lowest conformal dimension in the UV model (negative if the
model is non-unitary). Taking into account this fact, it is possible to
identify
the UV models with non-unitary $\frac{SU(2)_k\otimes SU(2)_l}{SU(2)_{k+l}}$
cosets, with $l$ integer and $k=\frac{p}{2}-2$. The perturbing operator with
$\Delta=\frac{p-2}{p+2}$ turns out to be the usual $\phi_{1,1,3}$ field. The
first series $l=1$ is given by the non-unitary minimal models $M_{p,p+2}$,
perturbed by their $\phi_{13}$ operator. The second series is supersymmetric,
the models are $SM_{p,p+4}$. One of these models also belongs to the minimal
series, namely the first $SM_{3,7}\equiv M_{7,12}$. This gives therefore also a
description of the flow of the theory $M_{7,12}+\phi_{4,5}$.

The S-matrices of these models are a tensor product of minimal Bernard LeCLair
S-matrices times Smirnov reductions~\cite{smirnov}
of the Sine-Gordon S-matrix for fractional
value of $k$
\eq
S=S_k \otimes S_l \virg l\in {\bf Z} \virg k \in {\bf Z}+\frac{1}{2}
\en
This also is in agreement with the ``gluing'' procedure suggested in~\cite{FI}.

\resection{Conclusions, generalizations and final remarks}

We have explored the whole class of magnonic TBA's whose Y-system is of the
form
(\ref{Y}). Of course, this is far from being the most general case.
In~\cite{wtba} it has been shown how the TBA's for higher coset models
perturbed
by $\phi_{id,id,adj}$ organize in a nice way in terms of two Dynkin diagrams,
one pertaining the physical particles (call it $G$) and one the magnons (call
it
$H$). The general TBA for coset models has the universal form
\eq
\nu_a^i = \varepsilon_a^i + \frac{1}{2\pi} \phi_g *
\left\{\sum_{b=1}^r G_{ab}[\nu_b^i-\log(1+e^{\varepsilon_b^i})]
- \sum_{j=1}^s H^{ij} \log(1+e^{-\varepsilon_a^j})\right\}
\label{gtba}
\en
where $g={\rm cox}\,G$, $r = {\rm rank}\,G$, $s={\rm rank}\,H$. We introduce
the
notation $G \diamond H$ for the graph encoding of this TBA. This ``product'' is
of course non-commutative, as one can not exchange the role of particles and
magnons in general. The graph alone is not sufficient to encode the TBA: one
still has to specify the form of the $\nu_a^i$. The rule is to encode masses
proportional to the Perron-Frobenius of $G$, for a single node in $H$, while
all
the other are zero (or, when possible, left and right movers in the usual way).
The diagonal TBA explored by Klassen and Melzer~\cite{KM} correspond to $G
\diamond A_1$ in this notation. For larger $H$ one has to indicate on which
node $k$ the mass terms must be put, we do that by adding an index $k$ to the
whole $G\diamond H$ symbol.
The magnonic TBA's studied in the present paper correspond
to $(A_1\diamond H)_k$, for all $k\in H$. What we have proved in sect.4 amounts
to the statement that, considering general TBA's of the form (\ref{gtba}), the
case $G\diamond A_1$ admits sensible solutions only for $G=A,D,E,T$ and
analogously $A_1 \diamond H$ allows only for $H=A,D,E,T$. Unfortunately, we
were
not able to find a similar classification for the general $G\diamond H$, in any
case the set of $G,H$ running on all $ADET$ is already extremely rich.
In~\cite{wtba} the case of $G=ADE$ and $H=A$ only has been explored. We expect
the other cases to hidden some beautiful surprise~\cite{prep}.

If, along the
same lines of sect.3, we search for the Y-system corresponding to the TBA
(\ref{gtba}), we get (here $Y_a^i=e^{\varepsilon_a^i}$)
\eq
Y_a^i\left(\theta +\frac{i\pi}{g}\right)
Y_a^i\left(\theta-\frac{i\pi}{g}\right)=
\prod_b (1+Y_b^i(\theta))^{G_{ab}} \prod_j (1+\frac{1}{Y_a^j(\theta)})^{H^{ij}}
\label{Ygen}
\en
This system shows a periodicity $Y_a^i(\theta+i\pi P)
=Y_{\bar{a}}^{\bar{\imath}}(\theta)$, with $P=\frac{h+g}{g}$, that encompasses
and generalizes both cases analyzed in this paper.

We would like to conclude by mentioning an intriguing observation.
The Y-system (\ref{Ygen}) shows
a curious duality: by exchanging $\varepsilon \to -\varepsilon$
we go from the Y-system of the $G\diamond H$ case to that of $H\diamond G$.
This clarifies why the two cases encompassed in the present paper had similar
Y-system, related exactly by a sign flip in the $\varepsilon$'s. More generally
one can think of some relation existing between the models described by TBA's
dual to each other.
We are at present not able to give any clear statement on this
subject, simply we notice the amusing fact that the series of tadpole diagrams
$T_n \diamond A_1$
considered by Klassen and Melzer to describe the perturbation of $M_{2,2n+3}$
minimal models by their $\phi_{13}$ field, goes under the described operation
into the $A_1 \diamond T_n$ case. Now, if we consider the $(A_1\diamond T_n)_1$
TBA, this describes the perturbation of $M_{2n+1,2n+3}$ by its $\phi_{13}$
field. What is surprising is that the two models $M_{2,2n+3}$ and $M_{2n+1,
2n+3}$ are known to be related by {\em level-rank duality}~\cite{Alt}. If this
important property of CFT has a relation to the duality described here or not,
has to be explored further~\cite{prep}.

\vskip .6cm
{\bf Acknowledgements} -- We are greatly indebted to F.Gliozzi for a lot of
very
useful discussions, especially on the $ADE$ classification of Y-systems. In
particular R.T. would like to thank F.Gliozzi for the patient and competent
tutoring during this first part of PhD in Torino. We also are grateful to
M.Alves, C.Destri, H.De Vega, P.Dorey, V.Fateev, A.Koubek, G.Mussardo
and I.Pesando for useful discussions and help.
R.T. thanks the Theory Group at Bologna University for the
kind hospitality during the final part of this work.

\end{document}